# Effect of Thermal Phase Fluctuations on the Superfluid Density of Two Dimensional Superconducting Films

Stefan J. Turneaure and Thomas R. Lemberger
*Dept. of Physics*
*Ohio State University*
*Columbus, OH 43210-1106*

John M. Graybeal
*Dept. of Physics*
*University of Florida*
*Gainesville, FL 32611-8440*

High precision measurements of the complex sheet conductivity of superconducting $Mo_{77}Ge_{23}$ thin films have been made from 0.4 K through $T_C$. A sharp drop in the inverse sheet inductance, $L^{-1}(T)$, is observed at a temperature, $T_C$, which lies below the mean-field transition temperature, $T_{C0}$. Just below $T_C$, the suppression of $L^{-1}(T)$ below its mean-field value indicates that longitudinal phase fluctuations have nearly their full classical amplitude, but they disappear rapidly as T decreases. We argue that there is a quantum crossover at about 0.94 $T_{C0}$, below which classical phase fluctuations are suppressed.

The nature of the superconducting to normal transition in two dimensional superconductors has long been of interest [1,2]. Prior studies of the complex conductivity below $T_C$ in two dimensional superconducting films and arrays have focused on the role of thermally generated vortex-antivortex pairs and the associated Kosterlitz Thouless Berezinskii (KTB) unbinding transition [3-5]. Largely ignored has been the suppressive effect of nonvortex, longitudinal current fluctuations on the superfluid density. Recently it has been suggested that longitudinal phase fluctuations could account for, or at least contribute to, the linear in temperature, T, suppression of the superfluid density, $n_s(T)$, [6-8] observed in the cuprates well below $T_C$ [9,10]. Experimentally, $n_s(T)$ is proportional to the inverse sheet inductance, $L^{-1}(T)$, of a two-dimensional (2-D) layer.

According to Ginzburg-Landau theory the superfluid density, $n_s(T)$, should be suppressed by a factor proportional to the mean square superfluid momentum [11]. A simple calculation utilizing the equipartition theorem for the kinetic energy of all Cooper pairs in a coherence volume, $\pi\xi^2(T)d$, where $\xi(T)$ is the Ginzburg-Landau coherence length and d is the film thickness, leads to the following result:

$$1 - n_s(T)/n_{s,MF}(T) \approx (2/\pi) k_B T/U_0(T). \tag{1}$$

$n_{s,MF}(T)$ is the mean-field superfluid density in the absence of phase fluctuations and $U_0(T) \equiv (\phi_0/2\pi)^2 L^{-1}(T) \approx \hbar\, 1.027\, k\Omega\, L^{-1}(T)$ is a characteristic superconducting energy. Equation (1) correctly describes the T-dependence of the effect of classical phase fluctuations on $L^{-1}(T)$, as calculated in Refs. 6-8. We note that these calculations differ by more than a factor of 10 on the prefactor, i.e., $2/\pi$ in Eq. (1). For a square array of Josephson junctions, the prefactor is ¼.

The preceding discussion neglects the effect of quantum mechanics, which is to suppress phase fluctuations below their classical size. In a single resistively-shunted Josephson junction, and in arrays of such junctions, quantum mechanics suppresses the noise currents that drive fluctuations in the phase difference across a junction when $k_BT/\hbar$ lies below a characteristic "R/L" frequency, $\omega_0$ [12]. The basic idea is that the magnitude of the noise current generated by the resistor is determined by the quantum Nyquist noise formula, and that low frequency noise currents pass through the inductive channel of the junction conductance while high frequency noise currents pass through the resistive channel. Phase fluctuations are reduced below their classical value by a factor,

$$f_Q \approx [1+ \hbar\omega_0 /k_BT]^{-1}, \qquad (2)$$

which is less than unity when $k_BT < \hbar\omega_0$.

The question arises as to whether similar considerations apply to continuous films. Following Ref. [12], for a continuous film, $\omega_0$ is roughly $1/GL$, where G is the quasiparticle sheet conductance, $\sigma_1 d$, averaged over frequencies up to $2\Delta_0(T)$ and wavevectors up to $2\pi/\xi$. Near $T_C$, $\sigma_1 d$ is only a little bit smaller than in the normal state, and we can approximate $1/G \approx R_n / [1 - L^{-1}(T)/L^{-1}(0)]$. For a dirty limit s-wave superconductor, like amorphous $Mo_{77}Ge_{23}$, T falls below $\hbar\omega_0/k_B$ when $T \approx .94\ T_{C0}$. Thus the model predicts that fluctuation effects should be small and nonlinear in T for $T < 0.9\ T_{C0}$. In prior studies this has been implicitly assumed without justification, inasmuch as $T_{C0}$ was determined by fitting $L^{-1}(T)$ with a polynomial over a temperature interval lying somewhat below $T_C$, and using the polynomial to extrapolate to $L^{-1}(T_{C0}) = 0$ [4]. In this letter we present data which justify this ad hoc procedure.

Although it was the high-$T_C$ cuprates which recently renewed interest in phase fluctuations we have chosen to study a-MoGe thin films for technical reasons. Isolating the effects of phase fluctuations in the high-$T_C$ cuprates is a difficult task since the mean-field inverse sheet inductance, $L^{-1}_{MF}$, is unknown and the measured $L^{-1}(T)$ generally varies between samples. On the other hand a-$Mo_{77}Ge_{23}$ thin films are nearly perfect weak coupling, dirty limit, s-wave superconductors for which the mean-field theory for $L^{-1}(T)$ is well known.

A series of 6 ultrathin $Mo_{77}Ge_{23}$ films were grown on 15mm diameter oxidized Si substrates by multitarget sputtering [13]. A thick film, labeled *A*, was grown on a Si substrate. It is 500Å thick and has the same $T_C$ as bulk material, and it serves as a control sample in which fluctuation effects are immeasurably small. The series of thin films, labeled *B* through *G*, range in thickness from 61Å to 21.5Å. Film properties are listed in Table I. $T_C$ for these films is defined as the temperature at which $L^{-1}(T)$ (defined below) has its maximum slope. $T_{C0}$ decreases as the film thickness decreases due to 2D localization effects [14] but the resistivity is nearly independent of film thickness. The sheet resistance at 15K can be well approximated by[14]

$$R_n = [16.7 \text{ k}\Omega\text{-Å}/d] \ (1 + 3\text{Å}/d). \tag{3}$$

The complex sheet conductance, $\sigma(\omega,T)d \equiv \sigma_1 d - i\sigma_2 d \equiv 1/[R + i\omega L]$, was measured using a two-coil mutual inductance technique with the drive and pick-up coils located on opposite sides of the film [15]. Data were taken continuously as T was ramped at less than 1 mK/s. Measurements were repeated at several frequencies ranging from

100 Hz to 50 kHz. The magnitude of the applied ac field was less than 20 µG at the center of the film for measurements near $T_C$, and the static magnetic field perpendicular to the film was reduced below 1 mG with a combination of µ-metal shielding and a 30 cm long field-nulling solenoid. Null magnetic field was found by maintaining a fixed sample temperature about 50 mK below $T_C$ and sweeping the magnetic field until the sheet impedance was minimized. Measurements were taken with several different drive current amplitudes to ensure that the film response was linear.

Except for very near $T_C$, $\sigma_2 \gg \sigma_1$, and the film's response is purely inductive. The data presented in this paper represent the highest frequency data, for which $L^{-1}$ is proportional to the "bare" superfluid density because the inductive impedance of thermally-excited vortices is much smaller than that of the superfluid [16]. We will consider the vortex contribution to $L^{-1}$ elsewhere.

The inset to Fig. 1 shows $\mu_0 L^{-1}(T)/d$ vs. T for all of the films. The systematic uncertainty in $L^{-1}(T)$ is about 4% due to uncertainty in the geometry of the coils. The normalized $L^{-1}(T)/L^{-1}(0)$ is known to very high precision, limited by temperature resolution and the stability of the electronics used to measure the pick-up coil voltage. The measured values for $L^{-1}(0)$ are all within 7% of the theoretical value, $L^{-1}_{th}(0) = \pi\Delta(0)/\hbar R_n$, which indicates good film quality. The energy gap, $\Delta(0)$, is determined from fits of the T dependence of $L^{-1}(T)/L^{-1}(0)$ to the theoretical form for dirty limit BCS superconductors [11]:

$$L^{-1}(T)/L^{-1}(0) = \delta(T) \tanh [ \delta(T) \Delta(0)/(2k_B T)], \qquad (4)$$

where $\delta(T)$ is $\Delta(T)/\Delta(0)$.

The main panel of Fig. 1 shows $L^{-1}(T/T_{C0})/L^{-1}(0)$ vs. $T/T_{C0}$ for $T/T_{C0} < 0.37$ for films A and E as well as the best fits to Eq. (4) (dashed lines) [18]. Note that the vertical scale is greatly expanded near unity. From fits to the data we find that $\Delta(0)/k_B T_{C0} =$ 1.89 (0.1) for all films [19] which is consistent with tunneling measurements [20] and with theoretical expectations [21]. The fit to the 500Å thick film data is perfect, but for the 30Å thick film there is a small extra T dependence below $0.2 T_{C0}$ which is not predicted by BCS theory. This deviation from BCS theory was observed in all of the films other than film A and grew systematically with decreasing film thickness.

Our first important result is the absence of classical phase fluctuations below $0.2 T_C$. The slope of the tangent to the data for film E at $0.1 T_{C0}$ is much smaller than the prediction of Eq. (1) (thick solid line). Furthermore, the deviation from BCS theory is better fit with an additive $T^2$ or $T^3$ term to the rhs of Eq. (4) rather than T-linear. The origin of this extra T dependence is not clear, but it is apparent that the data are inconsistent with classical phase fluctuations as described by Eq. (1). The question becomes, what is the quantum crossover temperature, above which classical fluctuations are restored?

Fig. 2 shows $L^{-1}(T/T_{C0})/L^{-1}(0)$ vs. $T/T_{C0}$ for all of the films. $T_{C0}$ was determined by fitting $L^{-1}(T)/L^{-1}(0)$ for each film to Eq. (4) between $0.7 T_{C0}$ and $0.85 T_{C0}$. The uncertainty in $T_{C0}$ is about 15% of $T_{C0}-T_C$. The gap ratio, $\Delta(0)/k_B T_{C0}$, was taken to be the same, 1.89, for all films. For these fits,

$$\delta(T) = [\cos(\pi/2\ T^2/T_{CO}^2)]^{1/2} \qquad (5)$$

was used [22]. It is remarkable that the data for all of the films collapse to a single curve with nearly the same value, slope and curvature between $0.7 T_{C0}$ and $0.85 T_{C0}$ with a single fitting parameter. If the quantum crossover had moved through the fitted temperature range as the sheet resistance of the films increased with decreasing thickness, the curves would not have overlapped so well. We conclude that the quantum crossover lies above 0.85 $T_{C0}$ for all films, and that the data for film *A* represent $L^{-1}_{MF}(T/T_{C0})/L^{-1}_{MF}(0)$ for all of the thin films.

The inset to Fig. 2 shows the transition region in more detail. All of the thin films exhibit a sharp drop in $L^{-1}(T)$ at $T_C$. Just below the drop, at $T = T_C^-$, $L^{-1}$ is 60 to 80% of its mean-field value. The drop occurs where $L(T)T/\mu_0$ has a value near 9.6mm-K, as predicted for the KTB transition, but this may be coincidental because the lower frequency dynamics, discussed elsewhere, are inconsistent with a KTB transition at $T_C$. As T decreases from $T_C$, $L^{-1}$ lies below its mean-field value, but it rises rapidly and equals its mean-field value within experimental resolution below 0.85 $T_{C0}$ [23]. This agrees with the prediction of the quantum crossover in a qualitative manner, with the crossover occurring near $T_C$.

A rigorous quantitative analysis is not possible given the available theory, but we can make an approximate analysis. Based on Eq. (1), we expect phase fluctuations to increase approximately proportional to $k_B T/U_{00}(T)$, where $U_{00}(T) \equiv (\phi_0/2\pi)^2 L^{-1}_{MF}(T)$. Figure 3 shows $L^{-1}/L^{-1}_{MF}$ vs. $k_B T/U_{00}(T)$ where $L^{-1}_{MF}(T)$ is from the 500Å thick film (Fig. 2). The thick solid line was obtained from Eq. (1) by replacing $2/\pi$ with ¼, then replacing $n_s(T)/n_{s,MF}(T)$ by $L^{-1}/L_{MF}^{-1}$ and solving for $L^{-1}/L_{MF}^{-1}$. This provides a self-consistent result for the fluctuation-suppressed inverse inductance. The replacement of $2/\pi$ with ¼ was chosen to agree with the classical result for a square lattice of Josephson junctions at low T, where fluctuations are small. Also, Minnhagen has suggested that longitudinal classical phase fluctuations suppress the inverse inductance $L^{-1}/L^{-1}_{MF}$ with a slope of -1/4 in the 2D-XY model [2] (dashed line in Fig. 3). We believe that the self consistent calculation, which includes a discontinuity in $L^{-1}$ when $L^{-1} = 1/2\, L^{-1}_{MF}$ and $k_B T/U_{00}(T)=1$, is a reasonable extension to temperatures where fluctuations are large.

Illustrating the rapid onset of fluctuations, the thin lines in Fig. 3 show $L^{-1}/L^{-1}_{MF}$ for five of the thin films. The inset shows that up to $k_B T/U_{00}(T) = 0.2$, $L^{-1} = L^{-1}_{MF}$ within experimental resolution for all films. Note that $k_B T/U_{00}(T) = 0.2$ corresponds to $T/T_{C0}$ between 0.76 (film G) and 0.86 (film B). Above $k_B T/U_{00}(T) = 0.4$, which corresponds to $T/T_{C0} > 0.88$ for all films, $L^{-1}/L^{-1}_{MF}$ abruptly curves below unity, as expected for a quantum crossover. Once the crossover from quantum to classical regimes begins it occurs very rapidly since a decrease in $L^{-1}$ decreases $\omega_0$ in Eq. (2), which implies a smaller quantum suppression, and thus a further decrease in $L^{-1}$. The abrupt drop in $L^{-1}/L^{-1}_{MF}$ occurs at $L^{-1}/L^{-1}_{MF} \approx 0.6$ and $k_B T/U_{00}(T) \approx 0.9 \pm 0.15$. Uncertainty in $T_{C0}$ for each film leads to uncertainty in $U_{00}$, and therefore in the value of $k_B T/U_{00}(T)$ at which $L^{-1}/L_{MF}^{-1}$ drops to zero. It is possible that the drops occur at a single value, e.g., $k_B T/U_{00}(T) \approx 1.0$.

In conclusion, we have shown that as far as the superfluid density is concerned, a-$Mo_{77}Ge_{23}$ thin films are nearly perfect BCS s-wave superconductors for which the gap ratio, $\Delta(0)/k_B T_{C0}$, is independent of thickness as predicted [21]. Below 0.2 $T_{C0}$, we observe weak power-law behavior which grows with increasing $R_n$, where BCS theory predicts that $n_s(T)$ is nearly constant. The effect is inconsistent with the predicted T-linear suppression due to classical longitudinal thermal phase fluctuations. In films thinner than about 100 Å, the effects of *longitudinal* phase fluctuations become noticeable when T is greater than about 0.9 $T_{C0}$, and then they increase rapidly. The data are qualitatively consistent with a quantum suppression of thermal phase fluctuations below 0.94 $T_{C0}$, as predicted for dirty-limit BCS superconductors in Ref. 12.

*Acknowledgments:* This work was supported in part by DoE grant DE-FG02-90ER45427 through the Midwest Superconductivity Consortium. One of the authors (S.J.T.) would like to acknowledge support from an Ohio State University fellowship.

TABLE CAPTION

I. Film Parameters. d is the nominal film thickness. $L^{-1}(0)$ is the measured inverse sheet inductance extrapolated to T = 0. The normal state sheet resistance, $R_n$, was calculated from Eq. (3). $T_C$ is defined as the temperature at which $L^{-1}(T)$ has its maximum slope. $T_{C0}$ was determined by fitting data between 0.70 $T_{C0}$ and 0.85 $T_{C0}$ to Eqs. (4) and (5), with $\Delta(0) = 1.89\ k_B T_{C0}$. The uncertainty in $T_{C0}$ is about 15% of $T_{C0} - T_C$.

Table I.

| Film | A | B | C | D | E | F | G |
|---|---|---|---|---|---|---|---|
| d (Å) | 499 | 61 | 46 | 37 | 30 | 27.5 | 21.5 |
| $L^{-1}(0)$ (nH)$^{-1}$ (±4%) | 162 | 13.2 | 9.55 | 6.70 | 4.97 | 4.21 | 2.57 |
| $R_n$ (Ω) (±5%) | 33.7 | 287 | 387 | 488 | 612 | 674 | 885 |
| $T_{C0}$ (K) | 7.050 | 5.559 | 5.040 | 4.640 | 4.149 | 3.881 | 3.167 |
| $T_C$ (K) (±5 mK) | 7.050 | 5.442 | 4.917 | 4.499 | 3.998 | 3.734 | 2.999 |
| $(T_{C0} - T_C)/T_{C0}$ (±15%) | 0 | 0.021 | 0.024 | 0.030 | 0.036 | 0.038 | 0.053 |

FIGURE CAPTIONS

1. $L^{-1}(T/T_{C0})/L^{-1}(0)$ vs. $T/T_{C0}$ for films *A* and *E*. The data are shifted vertically for clarity. The dashed lines are fits to the dirty limit BCS theory given by Eq. (4) [18]. For the 6 thinnest films, there is a slight deviation from theory, which grows as the film thickness decreases. For film *E* the deviation is less than 0.1% of the zero temperature value. The inset shows $\mu_0 L^{-1}(T)/d$ for all 7 films.

2. $L^{-1}(T/T_{C0})/L^{-1}(0)$ vs. $T/T_{C0}$ for all films except *E*. Below 0.85 $T_{C0}$, $L^{-1}(T/T_{C0})/L^{-1}(0)$ vs. $T/T_{C0}$ is nearly identical for all of the films. The inset details the rapid onset of phase fluctuations as T exceeds 0.85 $T_{C0}$ and the sharp drop in $L^{-1}(T)$ at $T_C$ for the thin films.

3. Measured suppression of the inverse sheet inductance below its mean-field value as a function of $k_B T/U_{00}(T)$ [thin lines]. The thick black line is the predicted suppression due to classical phase fluctuations [12]. The inset highlights data for $T/T_{C0} < 0.75$ [$k_B T/U_{00}(T) > 0.2$ corresponds to $T/T_{C0} > 0.75$ for all films; $k_B T/U_{00}(T) = 0.5$ corresponds to $T/T_{C0} > 0.9$.] $L^{-1}/L^{-1}_{MF}$ is unity except as T nears $T_C$, where the classical suppression is reached rapidly.

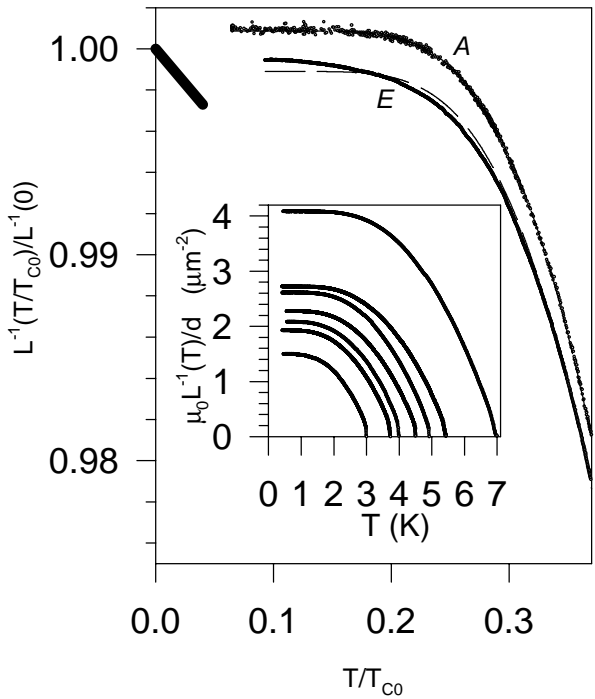

**Figure 1.**

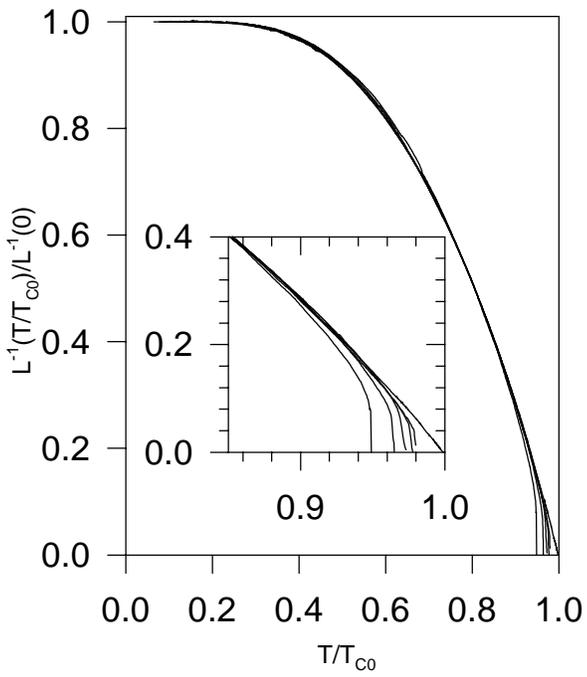

**Figure 2.**

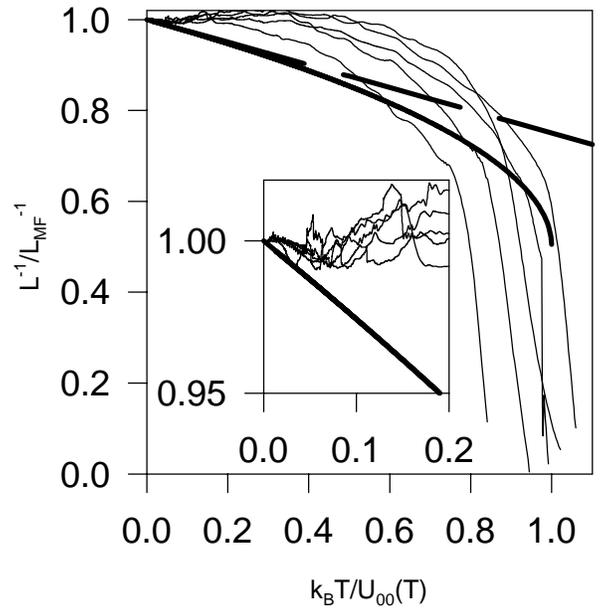

**Figure 3.**